\documentclass[superscriptaddress,amsmath,amssymb,aps,twocolumn,prx,notitlepage,10pt]{revtex4-2} 

\usepackage{amsmath,amsfonts,amssymb,amsthm,graphics,graphicx,epsfig,bbm} \usepackage[colorlinks=true,citecolor=blue,linkcolor=red,urlcolor=magenta]{hyperref} \usepackage[usenames]{color} \usepackage{graphicx} 
\usepackage{booktabs}
\usepackage{orcidlink} \usepackage{subfigure} \usepackage{amsmath} \usepackage{epsfig} \usepackage{dcolumn} \usepackage{bm} \usepackage{color} \usepackage{times} \usepackage{epstopdf} \usepackage{amssymb} \usepackage{amstext} \usepackage{latexsym} \usepackage{amsfonts} \usepackage{psfrag} \usepackage{soul,xcolor} \usepackage[normalem]{ulem} \usepackage{adjustbox} \renewcommand{\Re}{\mathrm{Re}}  \newcommand{\ket}[1]{\vert #1 \rangle}     \newcommand{\norm}[1]{\| #1 \|}

\begin{document}

\title{Regularized Counterdiabatic Driving for the Quantum Rabi Model}

\author{Juli\'an Ferreiro-V\'elez$^{\orcidlink{0000-0003-4864-7623}}$}
\email{julian.ferreiro@tecnalia.com}
\affiliation{TECNALIA, Basque Research and Technology Alliance (BRTA), 48160 Derio, Spain}
\affiliation{EHU Quantum Center and Department of Physical Chemistry, University of the Basque Country UPV/EHU, P.O. Box 644, 48080 Bilbao, Spain }

\author{Pablo Garc\'ia-Azor\'in$^{\orcidlink{0009-0001-9339-5210}}$}
\affiliation{EHU Quantum Center and Department of Physical Chemistry, University of the Basque Country UPV/EHU, P.O. Box 644, 48080 Bilbao, Spain }

\author{Francisco Andr\'es C\'ardenas-L\'opez$^{\orcidlink{0000-0002-2916-2826}}$}
\affiliation{Forschungszentrum J\"ulich, Institute of Quantum Control (PGI-8), D-52425 J\"ulich, Germany} 

\author{Xi Chen$^{\orcidlink{0000-0003-4221-4288}}$}
\email{xi.chen@csic.es}
\affiliation{Quantum Advanced Research Center (QuARC), CSIC, 28049, Madrid, Spain}
\affiliation{Instituto de Ciencia de Materiales de Madrid (ICMM), CSIC, 28049, Madrid, Spain}

\begin{abstract}

Counter-diabatic (CD) driving provides a powerful route to fast and robust state preparation by suppressing diabatic excitations during finite-time evolution. Yet, deriving analytical CD protocols for complex systems remains challenging, motivating the development of variational approaches. These methods typically rely on minimizing trace-based functionals to construct approximate control Hamiltonians. However, in unbounded systems, such functionals can become ill-defined because of the unbounded bosonic Hilbert space, leading to divergent cost functions and unphysical variational coefficients. Here, we introduce a variational optimization framework equipped with physically motivated renormalization schemes that regularize the trace-based metric by restricting it to relevant displaced and low-energy subspaces. As a paradigmatic example, we apply our method to the quantum Rabi model beyond the dispersive approximation and identify two distinct CD contributions that couple the atomic degree of freedom to the position and momentum quadratures of the field. These terms suppress diabatic excitations across coupling regimes ranging from strong to deep-strong light--matter interaction. We further formulate a fidelity-based quantum optimal-control strategy that bypasses the limitations of trace-based variational methods. Finally, we show that the resulting CD terms can be implemented via Floquet engineering through parametric modulation of the native Hamiltonian. Our results demonstrate that CD driving can be consistently extended to continuous-variable systems with unbounded Hilbert spaces, providing a controlled and scalable framework for quantum control in strongly interacting light-matter platforms.
\end{abstract}

\maketitle

\section{Introduction} 

The quantum Rabi model (QRM)~\cite{PhysRev.49.324} is one of the paradigmatic models of quantum optics and quantum information science, providing a minimal yet highly nontrivial description of light--matter interaction between a two-level system and a single bosonic mode. Since its original formulation and subsequent quantization~\cite{jaynes_original}, the QRM has served as a fundamental platform for exploring nonclassical state generation~\cite{q_info1,q_info2,experimental_cat}, parity-protected dynamics \cite{PhysRevLett.134.193604}, superradiant-like critical behavior~\cite{PT,PT2,Ani_PT,PT_yes2}, and strongly correlated spin--boson physics.

Beyond its theoretical relevance, the QRM has become a unifying framework for understanding and engineering a wide variety of quantum platforms, including cavity QED~\cite{qed_review}, trapped ions~\cite{trap_ions_review}, superconducting circuits~\cite{Wallraff_cqed,PhysRevA.80.032109,PhysRevLett.105.237001,Chiorescu2004,Niemczyk2010}, and periodically driven light--matter systems~\cite{PhysRevLett.134.063602}. In trapped ions, analog quantum simulation has enabled the QRM to be explored across a broad range of coupling regimes, including the strong- (SC), ultrastrong- (USC), and deep-strong-coupling (DSC) domains~\cite{RevModPhys.91.025005,USC_nori_review,PhysRevLett.105.263603,trapped_atoms_dsc,PhysRevX.8.021027}. In these regimes, the breakdown of the rotating-wave approximation (RWA), phonon wave-packet dynamics, and entangled ground-state features can be directly probed~\cite{PhysRevX.8.021027}.
In the USC and DSC regimes, counter-rotating terms become essential, excitation-number conservation breaks down, and the eigenstates acquire strongly nonclassical and highly entangled atom--field character. These regimes are of fundamental interest and provide new opportunities for quantum information processing~\cite{q_info_3,q_info_4} and quantum sensing~\cite{critical_sensing1,critical_sensing2}. 
However, preparing and manipulating QRM eigenstates across these regimes remains challenging. This difficulty is particularly pronounced in strongly hybridized and highly detuned regions, where small spectral gaps require slow adiabatic ramps that often exceed experimental coherence times \cite{PhysRevA.96.013849}. This motivates the development of fast and robust control for high-fidelity state preparation.

Shortcuts to adiabaticity (STA)~\cite{STA_review,STA_review2}, and in particular counterdiabatic (CD) driving ~\cite{Demirplak2005,Berry_2009,PhysRevLett.105.123003,PhysRevLett.111.100502,PhysRevLett.105.123003,PhysRevLett.111.100502}, provide a natural route to overcome these limitations. CD driving suppresses diabatic transitions by supplementing a reference Hamiltonian with an auxiliary control term generated by the adiabatic gauge potential (AGP)~\cite{jz_2013,geometry_Polkovnikov}, thereby reproducing adiabatic-like dynamics in finite time. STA techniques have already been applied to interacting spin--boson systems for fast and robust nonclassical-state engineering~\cite{PhysRevLett.124.180401}, and CD protocols have been proposed for the QRM in limiting cases such as the dispersive regime~\cite{Nori_sta}. Nevertheless, extending CD methods to the full QRM parameter landscape remains highly nontrivial. Exact CD constructions generally require complete spectral information and often generate experimentally inaccessible operator structures. Variational AGP methods~\cite{Claeys,sels_polkovnikov_minimizing} offer a practical alternative by optimizing an approximate control Hamiltonian within a restricted operator ansatz. However, these approaches are typically based on trace-based variational functionals over the full Hilbert space, which become problematic in continuous-variable systems. In the QRM, the bosonic mode introduces an unbounded Hilbert space, causing the standard trace-based variational action to depend pathologically on the Fock-space cutoff and, in the infinite-dimensional limit, to yield unphysical or vanishing CD coefficients.

In this work, we adopt variational and optimization techniques to extend previous dispersive-regime CD driving~\cite{Nori_sta} to arbitrary detuning ratios across the SC, USC, and DSC regimes and for different atom-field detunings. Building on variational AGP methods,  we first derive, in Sec. \ref{secII}, a first-order CD ansatz containing two physically distinct contributions: a photonic quadrature correction and an atomic quadrature correction. The photonic term reproduces the known dispersive-limit CD structure, while the atomic term becomes essential beyond the dispersive approximation and improves the driven-state fidelity in strongly hybridized regimes. We then show that the standard trace-based variational metric becomes ill-defined because the unbounded bosonic Hilbert space dominates the action. To overcome this limitation, in Sec. \ref{sec:renormalized_CD}, we introduce physically motivated renormalization schemes that restrict the variational metric to relevant displaced, low-energy, and symmetry-informed subspaces, thereby regularizing the optimization landscape and restoring meaningful AGP coefficients. As a complementary route, Sec. \ref{sec: variational_CD} reformulates the problem as a trace-free optimal-control strategy based on final-state fidelity, which bypasses the limitations of divergent trace functionals. Finally, Sec. \ref{sec: floquet} shows that the resulting CD interactions can be dynamically generated through periodic modulation of the native Hamiltonian, avoiding the need to introduce additional static coupling terms. 

Our results establish a controlled route for CD state preparation in continuous-variable light-matter systems. More broadly, they demonstrate that, in unbounded Hilbert spaces, the choice of variational metric is not merely a numerical detail but a physical ingredient of the AGP construction. This perspective provides a scalable framework for designing experimentally motivated CD protocols in strongly interacting bosonic and hybrid quantum platforms.



\section{Preliminaries}
\label{secII}

\subsection{Quantum Rabi model and its regimes}

\begin{figure*}[t]
    \centering
    \includegraphics[width=\textwidth]{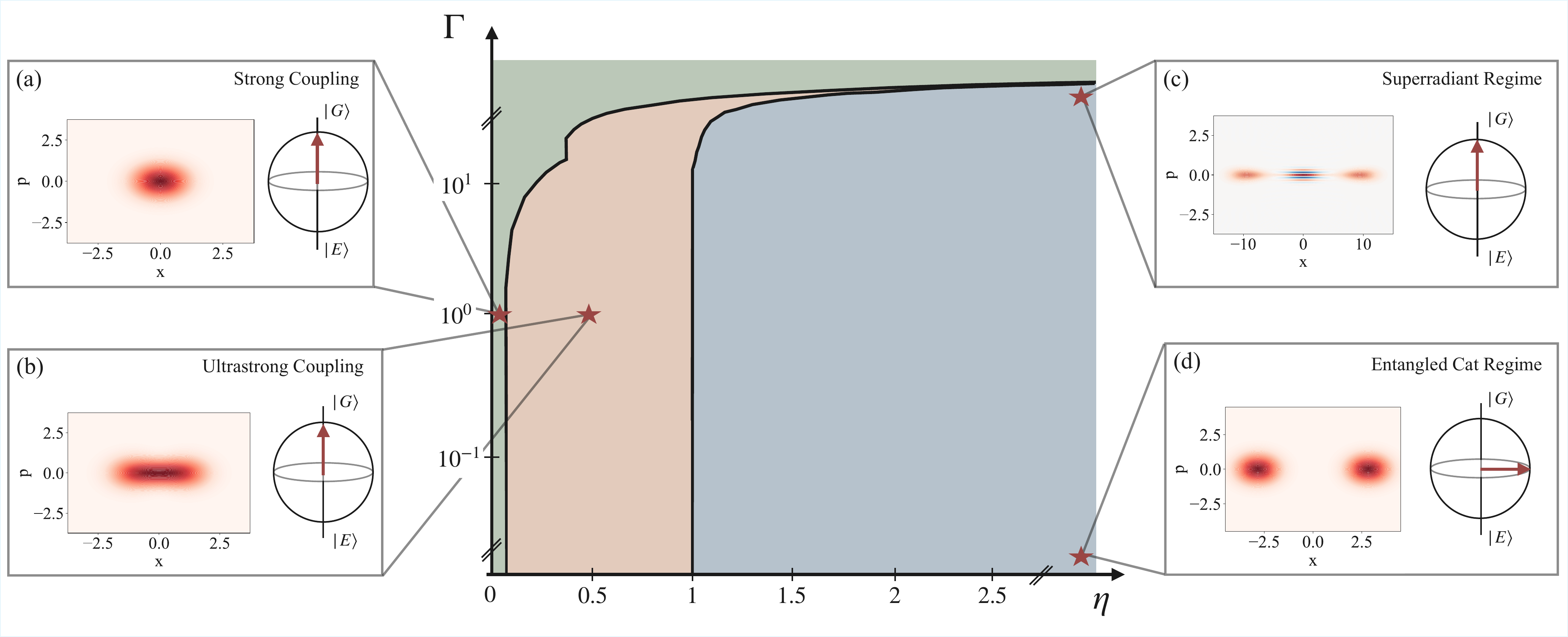}
    \caption{Regime map of the QRM across the normalized parameter space. The diagram divides the landscape into representative regions for different normalized couplings $\eta=g/\omega$ and frequency ratios $\Gamma=\Omega/\omega$, including the SC  (green), USC (red), and DSC  (blue) regimes. The SC-USC boundary is determined by the breakdown of the RWA, quantified by the fidelity condition $F(|\psi_{\rm JC}\rangle,|\psi_{\rm R}\rangle)\leq 0.9$. The USC-DSC boundary is defined by the onset of appreciable field occupation, $\langle \hat n\rangle\geq 1$, signaling the emergence of interaction-induced displacement. The adjacent panels display the Wigner function and the Bloch sphere for representative light-matter states: (a) SC regime, $\Gamma=1$ and $\eta=0.1$; (b) USC regime, $\Gamma=1$ and $\eta=0.4$; (c) entangled-cat regime, $\Gamma=10^{-4}$ and $\eta=1$; and (d) superradiant regime, $\Gamma=10^4$ and $\eta=1.005 \eta_c $, where $\eta_c = \sqrt{\Omega/4\omega}$  is the critical point \cite{PT}. The figure illustrates the qualitative changes in light-matter correlations across the different regimes.}
    \label{fig: sketch_rabi_regimes}
\end{figure*}

Our starting point is the quantum Rabi Hamiltonian ($\hbar \equiv 1 $) \cite{PhysRev.49.324,Braak_int}
\begin{equation}
\label{rabi_ham}
    \hat{\mathcal{H}}_R =  \omega \hat{a}^\dagger\hat{a} + \frac{ \Omega}{2}\hat{\sigma}_z +  g \hat{\sigma}_x (\hat{a}^\dagger+ \hat{a}).
\end{equation}
This model describes the interaction between a two-level system of transition frequency $\Omega$ and a single quantized bosonic mode of frequency $\omega$, coupled with strength $g$. Here, $\hat a$ and $\hat a^\dagger$ are the annihilation and creation operators of the bosonic mode, while $\hat\sigma_i$ denotes the $i$-th Pauli operator acting on the two-level system. Despite its simple form, the QRM exhibits a hierarchy of dynamical regimes~\cite{PhysRevA.96.013849,Dif_regims1,classification_2}, arising from the competition between the bare frequencies and the light--matter coupling strength, as illustrated in Fig.~\ref{fig: sketch_rabi_regimes}.
To compare the different regimes on equal footing, we work with the dimensionless Hamiltonian obtained by rescaling energies by $\omega$,
\begin{equation}
\label{rabi_ham_ren}
    \hat{\mathcal{H}}_R = \hat{a}^\dagger\hat{a} + \frac{\Gamma}{2}\hat{\sigma}_z +  \eta \hat{\sigma}_x (\hat{a}^\dagger+ \hat{a}),
\end{equation}
where $\Gamma=\Omega/\omega$ is the frequency ratio and $\eta=g/\omega$ is the normalized light-matter coupling. Throughout this work, we focus on the coherent dynamics generated by the closed-system Hamiltonian. Dissipative processes, such as spontaneous emission at rate $\gamma$ and photon loss at rate $\kappa$, are neglected under the assumption that $(\kappa,\gamma)\ll(\omega,\Omega,g)$ on the timescales of interest. This allows us to isolate the role of CD corrections across the different QRM regimes.

As illustrated in Fig.~\ref{fig: sketch_rabi_regimes}, for weak light–matter interactions  $\eta\ll1$, the system remains in the SC regime \cite{Scully1997}. In this limit, the Hamiltonian preserves the continuous $U(1)$ symmetry, and therefore the system conserves the excitation number. In the SC, the RWA is valid, and the dynamics are accurately captured by the Jaynes–Cummings Hamiltonian \cite{jaynes_descendants}. As the coupling strengths increase $0.1 \lesssim \eta \lesssim 1$, the system enters into the USC regime \cite{RevModPhys.91.025005}. In this regime, counter-rotating terms are non-negligible, the RWA breaks down, and the $U(1)$ symmetry is replaced by the discrete $\mathbb{Z}_{2}$ symmetry \cite{Forn2016,PhysRevA.94.012328}. In the USC, the conserved operator is $\mathcal{P}=\sigma^{z}(-1)^{\hat{a}^{\dag}\hat{a}}$, and the state space is divided into two orthogonal subspaces with even and odd parity. Finally, we found the DSC regime, with $\eta \gtrsim 1 $ \cite{PhysRevLett.105.263603}. Here, the expectation number operator $\langle \hat{n} \rangle$ is no longer approximately zero, and the cavity state is represented by a displaced state. The Fig.  \ref{fig: sketch_rabi_regimes}-(d) exemplifies the macroscopic occupation of the cavity, within a ground state described by a Sch\"{o}dinger entangled cat state i.e., 
$|\Psi_{\rm G}\rangle \simeq
\frac{1}{\sqrt{2}}
\left(
|+_x\rangle|\alpha\rangle
-
|-_x\rangle|-\alpha\rangle
\right)$,
where $|\pm_x\rangle$ denote eigenstates of $\hat\sigma_x$, and $\alpha$ is the interaction-induced displacement. Thus, by varying $\eta$ and $\Gamma$, the QRM provides access to a broad family of nontrivial ground states with strong atom-field correlations.

Accessing the USC and DSC regimes poses notable experimental challenges. Nevertheless, several works have found that linear \cite{drive_1} or nonlinear driving protocols \cite{drive_2,drive_3} can enhance the effective light-matter coupling. In such setting, the normalized coupling can be parametrically controlled as
\begin{equation}
    \eta_{{\rm{eff}}}= \lambda(t)\eta_{{\rm{max}}},
\end{equation}
allowing the system to be driven into the USC or the DSC. Here, $\lambda(t)$ describes the scheduling function and  $\eta_{max}$ is the maximum attainable normalized coupling. To ensure adiabaticity, the system must be evolved over long timescales. However, decoherence and experimental constraints favor short protocols to mitigate experimental noise and loss. In this work, we use variational CD techniques to extend previous dispersive-regime CD~\cite{Nori_sta} to arbitrary detuning ratios $\Gamma$ across the SC, USC, and DSC regimes, enabling high-fidelity ground-state preparation across the QRM parameter landscape.

\subsection{CD driving and variational AGP}

CD driving, or transitionless quantum driving, is a quantum control technique designed to suppress diabatic excitations by adding an auxiliary control term to a reference Hamiltonian~~\cite{Demirplak2005,Berry_2009,PhysRevLett.105.123003,PhysRevLett.111.100502,PhysRevLett.105.123003,PhysRevLett.111.100502}. The objective is to reproduce the outcome of an adiabatic evolution within a finite, and typically much shorter, evolution time. The construction of the CD term starts by introducing a time-dependent unitary transformation $\hat U(t)$ that maps a reference orthonormal basis $\mathcal S$ into the instantaneous adiabatic basis, $\mathcal D(t)=\hat U(t)\mathcal S$. In this rotating frame, the reference Hamiltonian $\hat{\mathcal H}_0(t)$ is diagonal by construction, and the Hamiltonian governing the dynamics becomes
\begin{equation}
    \hat{\tilde{\mathcal H}}(t)
    =
    \hat U(t)\hat{\mathcal H}_0(t)\hat U^{\dagger}(t)
    -
    i\hbar \hat U(t)\frac{\partial \hat U^{\dagger}(t)}{\partial t}.
\end{equation}
Here, the first term is the rotated reference Hamiltonian, whereas the second is the inertial term or adiabatic gauge potential responsible for diabatic transitions. Transitionless driving consists of engineering an auxiliary Hamiltonian $\hat{\mathcal H}_{\rm CD}(t)$ satisfying
\begin{equation}
\hat U(t)\hat{\mathcal H}_{\rm CD}(t)\hat U^\dagger(t)
=
i\hbar \hat U(t)\frac{\partial \hat U^\dagger(t)}{\partial t},
\end{equation}
so that it exactly cancels the inertial contribution in the adiabatic frame~\cite{Demirplak2005}. The total Hamiltonian then reads
\begin{equation}
    \hat{\mathcal{H}}(t) = \hat{\mathcal{H}}_0(t) + \hat{\mathcal{H}}_{CD}(t).
\end{equation}
and enforces evolution along the instantaneous eigenstates of $\hat{\mathcal H}_0(t)$ independently of the driving rate. 

The main limitation of CD driving lies in the complexity of the exact auxiliary term. Although formally exact, analytic CD constructions are generally difficult to extend to many-body or continuous-variable systems, since they require spectral information about the instantaneous eigenstates of the Hamiltonian. Moreover, the exact CD Hamiltonian often contains highly nonlocal or experimentally inaccessible operator structures. To circumvent these limitations, variational approaches to CD driving have been developed~\cite{sels_polkovnikov_minimizing}. Instead of constructing the exact CD operator, these methods choose an ansatz within a restricted operator space and optimize it to approximate the exact AGP.

Within this framework, the CD Hamiltonian is expressed in terms of the AGP $\hat{A}_\lambda$ associated with the control parameter $\lambda(t)$, which generates the nonadiabatic response to the finite control times, relating the infinitesimal displacement in the control space $\delta \lambda$ with the corresponding corrections in the Hilbert space \cite{jz_2013}. Consequently, the CD term can be written as  
\begin{equation}
\label{CD_AGP_eq}
    \hat{\mathcal{H}}_{CD}(t;\vec{x}) = \dot{\lambda}(t) \hat{A}_\lambda(\vec{x}),
\end{equation}
 where $\vec{x}$ denotes the set of free parameters to optimize.  Although finding an exact closed form for the AGP is generally not feasible, it can be approximated systematically using a nested-commutator expansion,
\begin{equation}
    \label{nested}
    \hat{\mathcal{A}}_\lambda^{(l)}(t; \vec{x}) = i \sum_{k=1}^l x_k(t) \underbrace{[\hat{\mathcal{H}}_0(t),[\hat{\mathcal{H}}_0(t),...[\hat{\mathcal{H}}_0(t),}_{2k-1}\partial_\lambda \hat{\mathcal{H}}_0(t)]]].
\end{equation}
This expansion describes the AGP in a series of  $k$-th nested commutators \cite{Claeys,Petiziol,krylov_STA}. Higher orders improve expressivity but at expenses of generating more complex, nonlocal control terms. The expansion can therefore be truncated at low order to retain only controls compatible with the hardware capabilities and connectivity.  Once the AGP structure is defined, the variational coefficients $\vec{x}$ are optimized through a minimization of the action functional \cite{geometry_Polkovnikov,sels_polkovnikov_minimizing},
 \begin{align}
\label{action}
S(t;\vec{x}) &= \mathrm{Tr}\!\left[\hat{G}^\dagger(t;\vec{x})\hat{G}(t;\vec{x})\right], \\[6pt]
\label{G_operator}
\hat{G}(t;\vec{x}) &= \partial_\lambda \hat{\mathcal{H}}_0(t) - i \left[ \hat{\mathcal{H}}_0(t),
\hat{\mathcal{A}}_\lambda^{(l)}(t;\vec{x}) \right].
\end{align}

\begin{figure}[t]
    \centering
    \includegraphics[width=\linewidth]{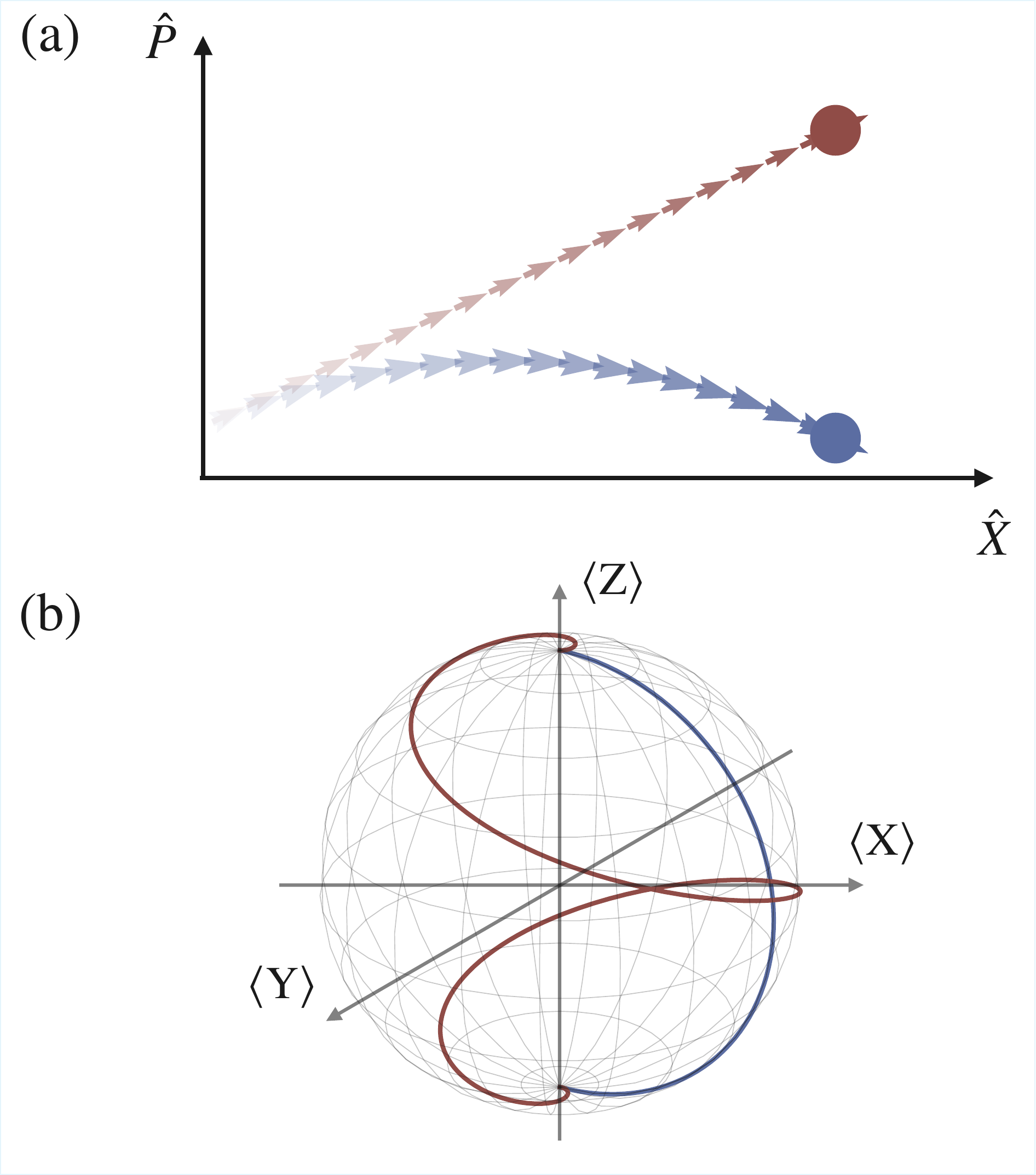}
  \caption{Schematic illustration of the CD corrections in the QRM. (a) Phase-space trajectories of the cavity quadratures. The red path corresponds to diabatic evolution without CD control, while the blue trajectory shows the corrected evolution produced by the cavity contribution $\hat{\mathcal{A}}_c = -i \eta \hat{\sigma}_x (\hat{a}^\dagger-\hat{a})$, which generates a spin-dependent force in the momentum quadrature. (b) Bloch-sphere representation of the qubit dynamics. The red trajectory shows the diabatic evolution, while the blue path shows the CD-corrected trajectory generated by the atomic contribution $\hat{\mathcal A}_a=\eta\Gamma\hat{\sigma}_y(\hat a^\dagger+\hat a)$.}
    \label{fig:STA}
\end{figure}

Next, we evaluate the first-order AGP correction for the time-dependent QRM, rewritten from Eq. (\ref{rabi_ham_ren}),
\begin{equation}
    \label{rabi_ham_time_dependent}
    \hat{\mathcal{H}}_R(t) = \hat{a}^\dagger \hat{a} + \frac{\Gamma}{2} \hat{\sigma}_z + \lambda(t) \eta \hat{\sigma}_x(\hat{a}^\dagger+\hat{a}). 
\end{equation}
with $\lambda(t) = \sin^2{[\pi\sin^2{(\pi t/(2 \tau}))}/2]$ being the scheduling function and $\tau$ the evolution time. By combining Eq.~(\ref{rabi_ham_time_dependent}) with the nested-commutator expansion in Eq.~(\ref{nested}), we can construct the first-order AGP ansatz,
\begin{equation}
\label{rabi_AGP}
    \hat{\mathcal{A}}_\lambda^{(1)} = x_1(t) \eta [ -i \hat{\sigma}_x (\hat{a}^\dagger-\hat{a})+ \Gamma \hat{\sigma}_y (\hat{a}^\dagger+\hat{a})]. 
\end{equation}
 Here, the first contribution,
$\hat{\mathcal{A}}_c=-i\eta\hat{\sigma}_x(\hat{a}^\dagger-\hat{a})$,
couples the spin degree of freedom to the momentum quadrature of the bosonic mode. This term compensates diabatic excitations associated with the interaction-induced displacement of the cavity field by correcting the motion along the conjugate quadrature, similarly to STA corrections in driven cavities~\cite{yin2022shortcuts}. The second contribution,
$\hat{\mathcal{A}}_a=\eta\Gamma\hat{\sigma}_y(\hat{a}^\dagger+\hat{a})$,
addresses diabatic errors originating from the interplay between the atomic term and the light--matter coupling. In particular, it corrects the qubit dynamics induced by the time-dependent interaction. Similar quadrature-mixing corrections have been reported in the optimal control of superconducting qubits through DRAG methods~\cite{DRAG}. Fig.~\ref{fig:STA} illustrates how the two CD terms compensate diabatic evolution in phase space and on the Bloch sphere for the field mode and the two-level system, respectively.

To determine the variational coefficient, one would normally minimize the action functional $S(t;\vec{x})$. However, in the QRM this procedure becomes ill-defined because the bosonic Hilbert space is unbounded. The trace functional receives contributions from increasingly highly excited Fock states, making the variational metric vanish proportionally to the Fock-space cutoff. This can be seen explicitly by evaluating the action in a truncated bosonic basis. For the first-order AGP ansatz in Eq.~\eqref{rabi_AGP}, the minimizing coefficient is given by
\begin{equation}
\label{analytic_agp_coef}
x_1(t) = \frac{-(1+\Gamma^2)}{1+\Gamma^4+6\Gamma^2+4\lambda^2(t)\eta^2[\frac{1}{n}+(2n-1)\Gamma^2]},
\end{equation}
where $n$ denotes the Hilbert space dimension of the field mode, implying the Fock-space cutoff. 
A detailed derivation is given in Appendix~\ref{Appendix_analytical_trace}. In the limit $n\rightarrow\infty$, the coefficient tends to zero for finite $\Gamma$, yielding a vanishing CD correction. This trivial result is not a physical prediction, but rather a consequence of applying a full trace norm to an unbounded Hilbert space. By contrast, in the dispersive limit $\Gamma\rightarrow0$, the atomic contribution in Eq.~\eqref{rabi_AGP} vanishes and the cutoff-dependent divergence disappears. In this limit, Eq.~\eqref{analytic_agp_coef} gives $x_1(t)\rightarrow -1$, and the CD Hamiltonian reduces to
\begin{equation}
\hat{\mathcal{H}}_{\rm CD}(t)
=
i\dot{\lambda}(t)\eta\hat{\sigma}_x(\hat{a}^\dagger-\hat{a}),
\end{equation}
which coincides with the CD ansatz previously obtained in the dispersive regime~\cite{Nori_sta}. This agreement validates the nested-commutator construction in the regime where the trace functional remains well defined. Beyond this limit, however, the cutoff dependence of Eq.~\eqref{analytic_agp_coef} shows that a regularized variational metric is required.

From this point onward, we therefore adopt a multiparameter formulation of the AGP, in which the cavity and atomic contributions are treated as independent variational components,
\begin{equation}
\hat{\mathcal{A}}_\lambda
=
\alpha_c(t)\hat{\mathcal{A}}_c
+
\alpha_a(t)\hat{\mathcal{A}}_a .
\end{equation}
This choice follows the broader variational CD strategy developed in quantum control and quantum computing ~\cite{PhysRevApplied.15.024038,PhysRevX.11.031070,VQE_STA,trott_dla}, where enlarging the physically motivated operator manifold can improve the approximation to the exact gauge potential and enhance the performance of finite-time protocols. In the present setting, independent variational weights allow the photonic and atomic diabatic channels to be corrected separately, increasing the expressibility of the CD protocol beyond the single-coefficient nested-commutator ansatz.

\section{Regularized variational metric}
\label{sec:renormalized_CD}

\begin{figure}[t]
    \centering
    \includegraphics[width=\linewidth]{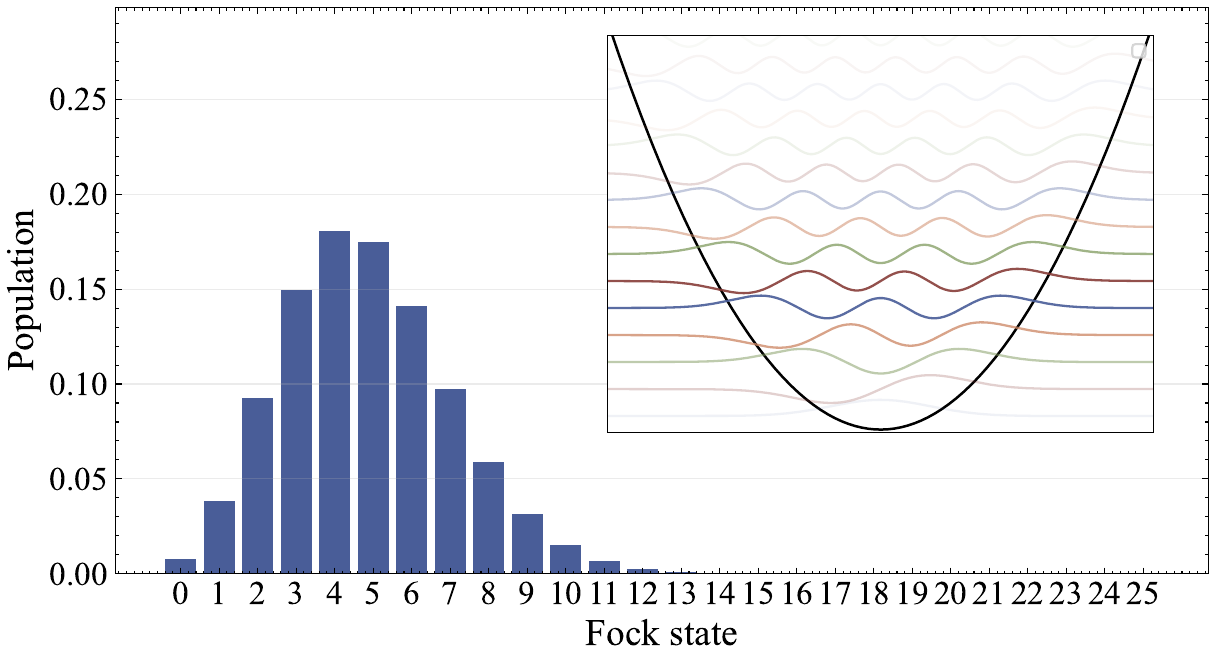}
    \caption{Schematic representation of the effective low-energy manifold of a cavity used to regularize the AGP optimization. The histogram represents the population of the renormalization state $(\rho)$ entering the weighted trace functional. The populated Fock states define the effective Hilbert-space region over which the variational metric is evaluated. }
    \label{fig: ho-thermalized}
\end{figure}

As discussed above, variational CD driving provides a powerful tool for approximating the AGP without requiring the full diagonalization of the Hamiltonian. The cost function defined in Eq.~\eqref{action} measures how well a variational ansatz satisfies the gauge-potential equation, or equivalently, the distance between the operator functional $\hat{G}(t;\vec{x})$ and the generalized force operator $\hat{M}_\lambda(t)$ generated by changes in the control parameter $\lambda(t)$. The exact AGP is recovered when $\hat{G}(t;\vec{x})=-\hat{M}_\lambda(t)$ is satisfied~\cite{jz_2013,geometry_Polkovnikov}.
It was already pointed out in the original variational-CD formulation that the full trace functional may be overly restrictive for determining approximate AGP coefficients, since it includes contributions from the entire Hamiltonian spectrum. Therefore, the optimization can be improved by restricting the suppression of diabatic transitions to a physically relevant subspace, such as a low-energy or low-temperature manifold, as schematically illustrated in Fig.~\ref{fig: ho-thermalized}. This leads to the regularized action
 \begin{equation}
 \label{renormalized_action}
    S_\rho(t;\vec{x})= \mbox{Tr}[\rho \hat{G}^\dagger(t;\vec{x})\hat{G}(t;\vec{x})],
\end{equation}
where $\rho$ is a reference density operator that restricts the variational metric to the relevant subspace. 

In this section, we explore physically motivated choices of the reference density operator $\rho$ within variational CD protocols. Our objective is not merely to reduce the computational cost of the optimization, but to incorporate physical prior information into the variational metric. By weighting the trace over the displaced and low-energy regions explored by the dynamics, this approach regularizes the action functional and removes the unphysical divergences associated with the unbounded bosonic mode of the QRM.

The reference operator $\rho$ can be chosen to encode physical information about the region of Hilbert space explored during the driven evolution. A natural choice for the bosonic sector in the QRM is a displaced thermal state,
\begin{equation}
    \label{displaces_thermal_state}
    \rho_c (\alpha,\beta) = \frac{1}{\mathcal{Z}} \hat{\mathcal{D}} (\alpha) e^{-\hbar\beta \omega \hat{a}^\dagger\hat{a}} \hat{\mathcal{D}}^\dagger(\alpha),
\end{equation}
where $\hat{\mathcal D}(\alpha)=\exp(\alpha\hat a^\dagger-\alpha^*\hat a)$ is the displacement operator, $\beta=(k_B T)^{-1}$, and $\mathcal Z=\mathrm{Tr}[e^{-\beta\hbar\omega \hat a^\dagger\hat a}]$ is the harmonic-oscillator partition function. This reference state incorporates both thermal weighting and the interaction-induced displacement of the bosonic mode.

Here, we focus on the coherent unitary dynamics of the ground state of the QRM and neglect dissipative effects. We therefore take the zero-temperature limit of Eq.~\eqref{displaces_thermal_state}, for which the bosonic reference state reduces to a displaced vacuum density matrix
\begin{equation}
     \rho_c (\alpha,\infty) = \hat{\mathcal{D}} (\alpha) |0\rangle \langle 0 |\hat{\mathcal{D}}^\dagger(\alpha) =  |\alpha\rangle \langle \alpha |.
\end{equation}
To estimate the displacement value in the renormalized action, we consider the limit of negligible atomic contribution, i.e. the dispersive regime
\begin{equation}
    \hat{\mathcal{H}}_{\Omega\rightarrow0}(t)
    = \omega \hat{a}^\dagger \hat{a}
    + \lambda(t) g\,\hat{\sigma}_x(\hat{a}^\dagger+\hat{a}) .
\end{equation}
In the eigenbasis of $\hat\sigma_x$, the bosonic Hamiltonian reduces to two displaced harmonic oscillators, which can be solved exactly as displaced Fock states, 
\begin{equation}
    \label{displaced_state}
    |\phi_\pm(t) \rangle = e^{\pm \lambda(t) \frac{g}{\omega}(\hat{a}^\dagger-\hat{a})} |n\rangle ,
\end{equation}
where the sign $(\pm)$ is fixed by the corresponding $\hat{\sigma}_x$ eigenvalue. Guided by this exactly solvable limit, we use the displacement scale $|\alpha(t)|=\lambda(t)g/\omega$ to construct the reference operator entering the regularized trace.
In our implementation, the qubit sector is weighted uniformly, while the cavity sector is centered around the two displaced branches. The corresponding coherent reference operator is
\begin{equation}
\label{renormalization_density_matrix}
    \rho_{coh}(t) =
    \frac{1}{4}\left[\mathbb{I}
    \otimes
    \left(\rho_c\left(\alpha,\infty\right)+\rho_c\left(-\alpha,\infty\right)\right)\right],
\end{equation}
where $\mathbb{I}= (|\uparrow \rangle \langle \uparrow | + |\downarrow \rangle \langle \downarrow |)$ explores both possible states of the qubit. The renormalized density matrix centers the variational metric around the displaced low-energy manifold relevant to the driven QRM dynamics.

The weighted trace in Eq.~\eqref{renormalized_action} provides a direct generalization of the trace functional by incorporating the regularization explicitly into the variational metric. This choice, however, is not unique. We therefore investigate two related regularization schemes and compare their performance.
First, we replace the Frobenius-type action used to evaluate the operator $\hat G(t;\vec{x})$ by a finite-temperature norm \cite{geometry_Polkovnikov},
\begin{equation}
    S_l(t;\vec{x}) = \langle \hat{G}^2(t;\vec{x})\rangle_{\rho_{\rm{ren}}} - \langle \hat{G}(t;\vec{x})\rangle_{\rho_{\rm{ren}}}^2.
\end{equation}
Here, the angular brackets denote the expectation value with respect to the reference state $\rho_{\rm{ren}} = |\psi_{\rm{sr}} \rangle \langle \psi_{\rm{sr}} |$, where the superradiant state is defined as \cite{sweeprabi}
\begin{equation}
|\Psi_{\rm{sr}} \rangle = \frac{1}{\sqrt{2}} (| \uparrow \rangle \otimes |\alpha \rangle + |\downarrow \rangle \otimes |-\alpha \rangle),
\end{equation}
and $\alpha$ is the displacement parameter derived above. Compared with the displaced-vacuum reference state in Eq. (\ref{renormalization_density_matrix}), this superradiant state explicitly incorporates atom-field entanglement.  The variance-based metric evaluated with the superradiant reference state allows us to assess whether including atom-field entanglement in the regularized action improves the variational protocol.

\begin{figure}[t]
    \centering
    \includegraphics[width=\linewidth]{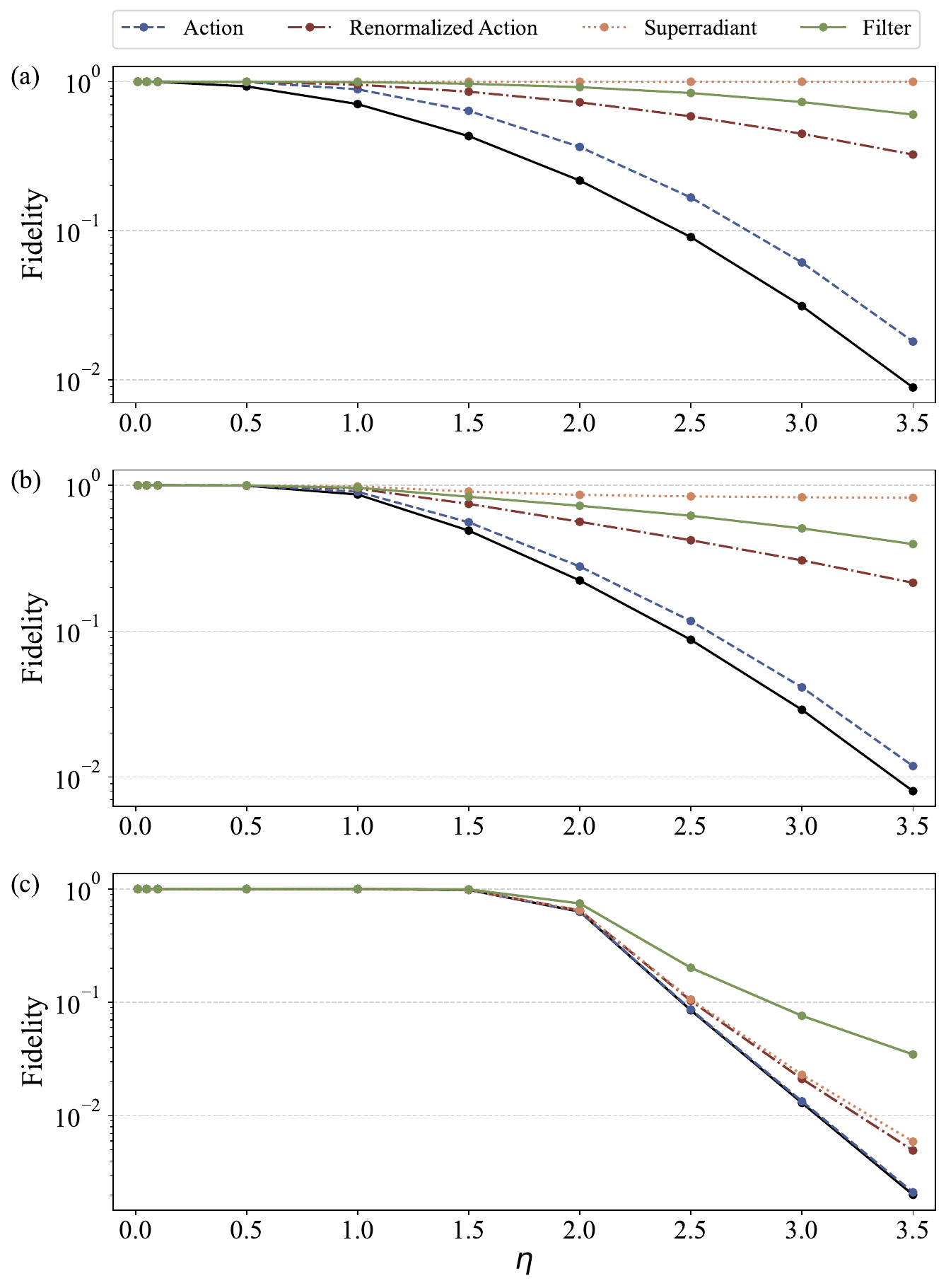}
    \caption{Final ground-state fidelity versus normalized coupling strength $\eta$, comparing the different renormalization schemes with the CD-free evolution (black solid line).  Results are shown for (a) the high-frequency cavity regime ($\Gamma=0.1$), (b) the resonance ($\Gamma=1$), and (c) the high-frequency atom regime ($\Gamma=10$). The simulations are performed at final time evolution $ \omega \tau =1$.}
    \label{fig: traces_fidelity}
\end{figure}

Next, we introduce a filtered-trace action, in which the reference operator is defined through an indicator function,
\begin{equation}
    \label{filter}
    \rho^{i,j}_f = \frac{1}{\mathcal{N}}\left(\mathbf{1}_{\rho_c^{ij}>\gamma}\right).
\end{equation}
Here $\mathcal{N}$ is a normalization factor and $\rho_c^{i,j}$ is the $(i,j)$ matrix element of the displaced vacuum state defined above, and $\gamma$ is the threshold parameter, which indicates if the density matrix $\rho_f$ is zero or the unity. The filter renormalization, unlike the previous weighted-trace schemes, enforces the convergence not through functional regularization but via domain restriction. The filter function, analogously to Schwartz compact supports, preserves the value of observables while inducing convergence through the truncation of the operator basis, suppressing contributions from highly excited states that are irrelevant to the driven dynamics.

We numerically minimize the corresponding regularized actions $S(t;\vec{x})$ and use the resulting coefficients to construct the CD terms. Fig.~\ref{fig: traces_fidelity} compares the final ground-state fidelity 
\begin{equation}
F(\rho, \sigma) = {\rm{Tr}} \left(\sqrt{\sqrt{\rho} \sigma \sqrt{\rho}}\right)^2,
\end{equation}
obtained with the different regularization schemes and with the CD-free evolution, across various detuning and interaction regimes.  The simulations are conducted in three representative configurations: the high-frequency cavity regime [Fig.  \ref{fig: traces_fidelity}(a)], the resonant regime [Fig.  \ref{fig: traces_fidelity}(b)], and the high-frequency atom regime [Fig.  \ref{fig: traces_fidelity}(c)], each explored at different coupling strengths.

The results indicate that incorporating a physically motivated reference state  into the action functional significantly improves  the performance of the variational AGP optimization. The regularized metrics lead to a systematic increase in the final ground-state fidelity across all the considered regimes, with the enhancement being particularly pronounced at stronger coupling. Fig.~\ref{fig: traces_fidelity} shows that  both the coherent-state weighted trace and the filtered trace improve the CD-assisted dynamics relative to the CD-free evolution. The superradiant state, which explicitly includes atom-field correlations, provides a noticeable additional improvement in the high-frequency-cavity and resonant regimes. In the high-frequency-atom regime, where the cavity mode becomes more strongly populated, the more restrictive nature of the filtered metric is beneficial for the AGP optimization.
A detailed analysis of the role of the reference state is presented in Appendix~\ref{app:variational_trace}, where we examine its influence on the topology of the action functional and on the statistical correlation between the action and the final fidelity. Interestingly, the results further reveal that regularization not only removes the cutoff-induced divergences associated with the bosonic mode, but also reshapes the gradient landscape of the functional, mitigating the emergence of barren-plateau-like regions.

Finally, we quantify the statistical relationship between the minimized action and the resulting fidelity by computing the Spearman's rank correlation coefficient (see Appendix~\ref{app:variational_trace}).
The results, summarized in Table \ref{spearsman_table}, demonstrate that incorporating physical prior information  into the variational metric substantially improves the predictive power of the action functional. 
This correlation indicates that the observed fidelity enhancement is not a stochastic artifact, but rather a consequence of restricting the variational search to physically relevant regions of Hilbert space.

\begin{table}[h]
\centering
\renewcommand{\arraystretch}{1.25}
\setlength{\tabcolsep}{6pt}
\begin{tabular}{c|c|c|c|c}
\hline
$(\omega,\Omega,g)$ & Action & R-Action & Superradiant & Filter \\
\hline
$(1,1,0.25)$ & -0.33  & -0.56  & -0.91  & -0.78 \\
$(1,1,0.8)$  & -0.15  & -0.31  & -0.37  & -0.21 \\
$(1,0.1,1)$  & -0.30  & -0.87  & -0.97  & -0.87 \\
$(1,10,1)$   & -0.18& -0.18& -0.18& -0.18\\
\hline
\end{tabular}
\caption{Spearman rank correlation coefficients $r_s$ between the accumulated action and the final ground-state fidelity for the different parameter regimes $(\omega,\Omega, g)$. The $r_s$ parameter represent the monotonic relationship between both quantities, demonstrating that the inclusion of renormalization schemes significantly enhances the predictive power of the action functional.}
\label{spearsman_table}
\end{table}

\begin{figure*}[t]
    \centering
    \includegraphics[width=\textwidth]{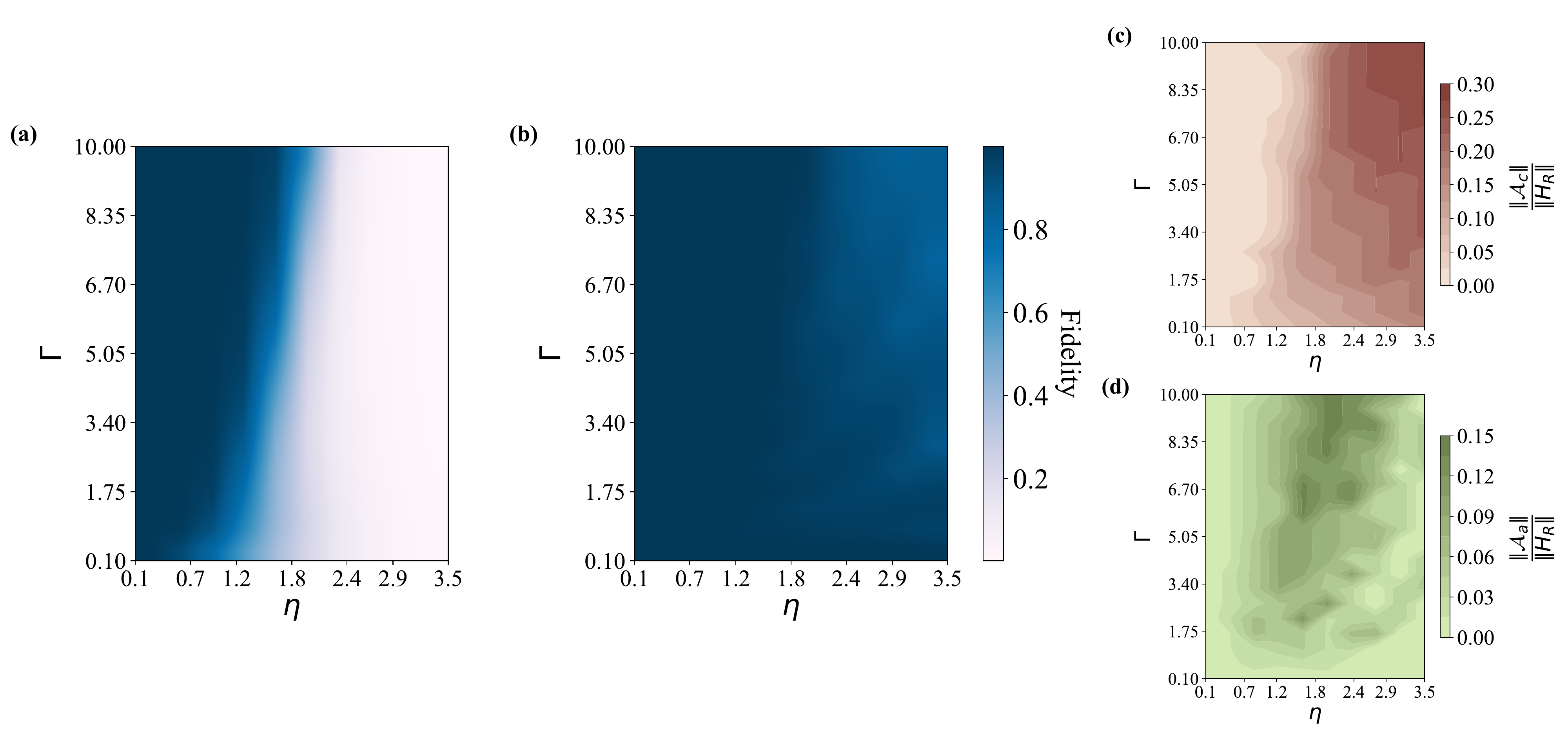}
    \caption{Simulated CD dynamics across the Rabi parameter space a dimensionless evolution time $\omega\tau=1$. (a) Ground-state fidelity of the unassisted Hamiltonian as a function of the normalized coupling strength $\eta = g/\omega$ and the frequency ratio $\Gamma = \Omega/\omega$. The fidelity landscape exhibits a pronounced degradation in the strong-coupling and highly detuned regimes, reflecting diabatic excitations. (b) Ground-state fidelity under CD driving using the optimized AGP. The assisted protocol restores high-fidelity transport across the entire parameter domain, substantially suppressing the diabatic errors that dominate the unassisted dynamics. (c,d) Relative norms of the optimized atomic and photonic AGP components, respectively,  showing the distinct dynamical contribution of the counterdiabatic corrections. These panels underscore how the counterdiabatic correction suppresses diabatic errors, and how the atomic and photonic AGP components compensate for them in distinct regions of the parameter space. }
    \label{rabi_regimes}
\end{figure*}



\section{Optimal AGP }
\label{sec: variational_CD}

In the previous section, we found that increasing the light-matter coupling can strongly enhance the cavity photon population. Accurately describing this regime requires increasing the bosonic Fock-space cutoff, which in turn makes the full trace action increasingly ill-conditioned. Although the regularized metrics introduced above mitigate this problem, it is useful to develop a complementary strategy that avoids trace-based functionals altogether.

To this end, we formulate a fidelity-based quantum optimal-control approach to determine the AGP coefficients. The goal is to construct CD protocols that remain effective across broad frequency and coupling regimes without relying on the full trace action. Similar difficulties arise in other CD settings, for instance near quantum phase transitions, where small spectral gaps make trace-based constructions challenging. Several alternative strategies have been proposed to address these issues, including efficient counterdiabatic paths \cite{efficient_paths_cd}, genetic optimization schemes \cite{genetic_cd}, mean-field counterdiabatic driving \cite{mean_cd}, and polynomial counterdiabatic expansions \cite{universal_prl,universal_prx}. Inspired by these optimization-based approaches, we determine the AGP parameters by maximizing a final-state fidelity through the following cost function
\small
\begin{equation}
\label{Cost_function}
    C(\vec{x}) = \left| \langle \Psi_{0}(\tau) | \exp{\left(-i\int_0^\tau  \hat{H}_R(t) + \dot{\lambda}(t) \mathcal{A}_\lambda(\vec{x})dt\right)} | \Psi_{0}(0) \rangle \right|^2,
\end{equation}
\normalsize
where $\ket{\Psi_0(0)}$ is the analytic ground state at $t=0$ and $\ket{\Psi_0(\tau)}$  is the ground state, obtained numerically in the truncated bosonic Hilbert space. 
In this approach, the AGP coefficients $\vec{x}$ are treated as time-independent variational parameters optimized to maximize the overlap between the evolved state and the target final ground state.

To preserve the physical interpretation of the CD correction, we constrain the optimization so that the AGP does not artificially replace the native Hamiltonian dynamics. By construction, the AGP itself is independent of the control velocity $\dot{\lambda}(t)$; the protocol duration enters only through the CD Hamiltonian $\hat{\mathcal H}_{\rm CD}(t)=\dot{\lambda}(t)\hat{\mathcal A}_\lambda$. Consequently, the CD contribution becomes negligible for slow protocols and dominant only in the impulse regime. Without an additional scale constraint, a direct fidelity optimization could generate unrealistically large AGP coefficients that improve the final overlap but no longer represent a physically meaningful counterdiabatic correction. We therefore require the AGP norm to scale with the natural light--matter coupling scale,
\begin{equation}
    \norm{\hat{\mathcal A}_\lambda}
    =
    \mathcal O(\eta),
\end{equation}
where $\norm{\cdot}$ denotes the operator norm, and $\eta$ is the normalized light--matter coupling. This constraint is consistent with the analytical scaling of the first-order AGP in Eq.~\eqref{analytic_agp_coef} and prevents the optimized CD term from overwhelming the native QRM dynamics.


\begin{figure}[t]
    \centering
    \includegraphics[width=1.0\linewidth]{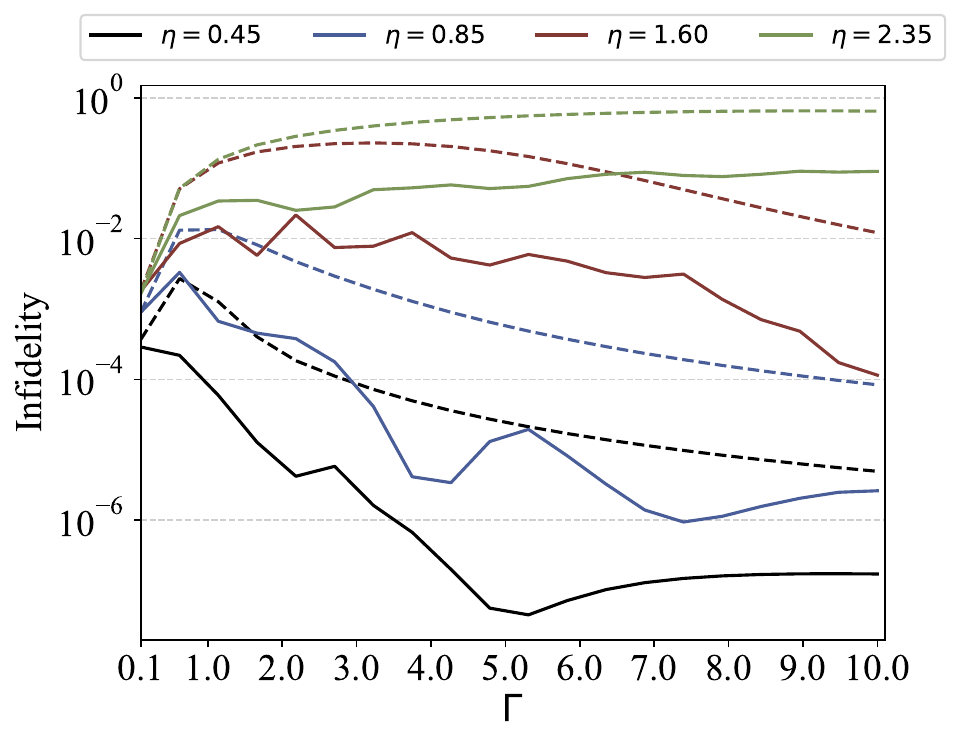}
    \caption{Infidelity of the analytically derived CD protocol (dashed lines) and the variationally optimized CD protocol (solid line), evaluated across a range of coupling strengths and frequency ratios. The optimized AGP achieves uniformly lower infidelity, typically by an order of magnitude, demonstrating the robustness of the optimized protocol.}
    \label{fig: nori_vs_grid}
\end{figure}

Figures~\ref{rabi_regimes}(a,b) compare the final ground-state fidelity of the unassisted evolution and the optimized CD protocol over a broad range of frequency ratios $\Gamma$ and normalized coupling strengths $\eta$. The unassisted dynamics exhibits pronounced fidelity loss in strongly coupled and highly detuned regimes, where diabatic excitations become significant. By contrast, the optimized CD protocol substantially improves the final fidelity throughout the explored parameter region. Figs.~\ref{rabi_regimes}(c,d) show the relative norms of the optimized atomic and photonic AGP components, normalized by the norm of the QRM Hamiltonian, $\norm{\hat{\mathcal A}_{a,c}}/\norm{\hat H_R}$.
These panels reveal the complementary roles of the two CD channels. The atomic correction is most relevant in regimes where the atomic frequency strongly affects the dynamics, while the photonic correction dominates when the cavity displacement induced by the light-matter coupling is the leading diabatic mechanism.

We further compare the optimized CD protocol with the analytical CD expression obtained in the dispersive regime from Eq.~\eqref{analytic_agp_coef}. Fig.~\ref{fig: nori_vs_grid} shows the resulting infidelity,
\begin{equation}
    I(\rho,\sigma) =  1-F(\rho,\sigma),
\end{equation}
as a function of the frequency ratio for several coupling strengths. The optimized AGP protocol systematically outperforms the dispersive analytical construction across the considered regimes, reducing the infidelity by approximately one order of magnitude. These results confirm that the additional atomic CD contribution provides a quantitative improvement in ground-state preparation beyond the dispersive approximation. They also highlight the robustness of the trace-free AGP optimization, which remains effective in strongly hybridized and highly detuned regimes.



\section{Perspectives on experimental implementation}
\label{sec: floquet}

In our theoretical proposal, the QRM Hamiltonian is supplemented by two CD contributions defined in the Eq.~(\ref{rabi_AGP}). While the native light-matter interaction term, proportional to $\propto \hat{\sigma}_x (\hat{a}^\dagger+a)$, couples the atomic degree of freedom to the position quadrature of the field, the additional CD involves cross-quadrature couplings. In particular, they couple the atomic and field quadratures in combinations that are not generally present in standard realizations of the QRM and may therefore be challenging to implement directly. To overcome this difficulty, we propose a Floquet-engineering strategy that dynamically generates the required operator structure without introducing additional control terms \cite{universal_floq,floquet_qs}. The central idea is to exploit high-frequency periodic modulations to synthesize an effective Hamiltonian whose dynamics reproduces the desired CD corrections \cite{Claeys,polarization_floquet_cd}. The Floquet theory states that a periodically driven Hamiltonian $\hat{\mathcal{H}}_E(t) = \hat{\mathcal{H}}_E(t+T) $ can be mapped onto an effective time-independent Hamiltonian $\hat{\mathcal{H}}_F $ through 
\begin{equation}
\label{floquet_stroboscopic_Hams}
    \exp \left( -i \hat{\mathcal{H}}_F T\right) \equiv \hat{\mathcal{T}} \exp \left( -i \int_t^{t+T} \hat{\mathcal{H}}_E(\tau) d\tau \right),
\end{equation}
where $T = 2\pi/\nu$ is the driving period and $\hat{\mathcal{T}}$ denotes the time ordering. In the high-frequency regime, the effective Hamiltonian can be  derived systematically using the Magnus expansion. This approximation formulates the effective Hamiltonian as a controlled series of nested commutators in powers of $\nu^{-1}$ (see Appendix \ref{floquet}) that allows to engineer approximate AGP into the evolution. In order to achieve the desired structure, we define the stroboscopic Hamiltonian as
\begin{align}
\label{eq:oscilating_floquet_rabi}
        \hat{\mathcal{H}}_E(t) &=   \hat{\mathcal{H}}_R(t) + \frac{\nu}{\nu_0}\cos(\nu t) \left[ A_c(t)  \hat{a}^\dagger\hat{a} +   A_a(t)   \hat{\sigma}_z\right] \nonumber \\[6pt]
   & \quad +  \dot{\lambda}(t) f^{(k)}(t) \eta \hat{\sigma}_x (\hat{a}^\dagger+ \hat{a}),
\end{align}
with
\begin{equation}
\label{oscilation_function}
f^{(k)}(t) = \sum_{k=1}^\infty \beta_k \sin((2k-1)\nu t).
\end{equation}
where $\nu_0$ is a reference frequency scale, $A_c(t)$ and $A_a(t)$ are slowly varying cavity and atomic modulation envelopes, respectively, and $\beta_k$ are Fourier coefficients defining the interaction-modulation profile. The additional modulation functions $A_c(t)/A_a(t)$  provide an independent degree of freedom for shaping the effective AGP operators generated by the nested-commutator independently. Restricting the Magnus expansion to leading order and retaining only the first harmonic of the interaction modulation, the stroboscopic effective Hamiltonian, up to the order $\mathcal{O}(\nu^{-2})$, becomes
\small
\begin{eqnarray}
\label{eq:effective_floquet_ham}
    \hat{\mathcal{H}}_F = \hat{\mathcal{H}}_R+ \dot{\lambda}(t) \left[ \frac{\eta \beta \bar{A}_a}{2 \nu_0} \hat{\sigma}_y (\hat{a}^\dagger + \hat{a}) + \frac{i \eta\beta \bar{A}_c}{4 \nu_0} \hat{\sigma}_x (\hat{a}^\dagger-\hat{a})\right],
\end{eqnarray}
\normalsize
where $\bar{A}_c$ and  $\bar{A}_a$ are the envelope-averaged amplitudes. Importantly, the emergent interaction terms reproduce exactly the operator structure of the first-order adiabatic gauge potential derived in Eq.~(\ref{rabi_AGP}). By appropriately tuning the modulation amplitudes $A_c(t)$ and $A_a(t)$, the coefficients of the effective Hamiltonian can be matched to the optimal variational parameters obtained in the previous sections. In this way, the desired counterdiabatic trajectory is realized dynamically through periodic driving, without the need to introduce additional static couplings into the Hamiltonian. 

\begin{figure}[t]
    \centering
    \includegraphics[width=\linewidth]{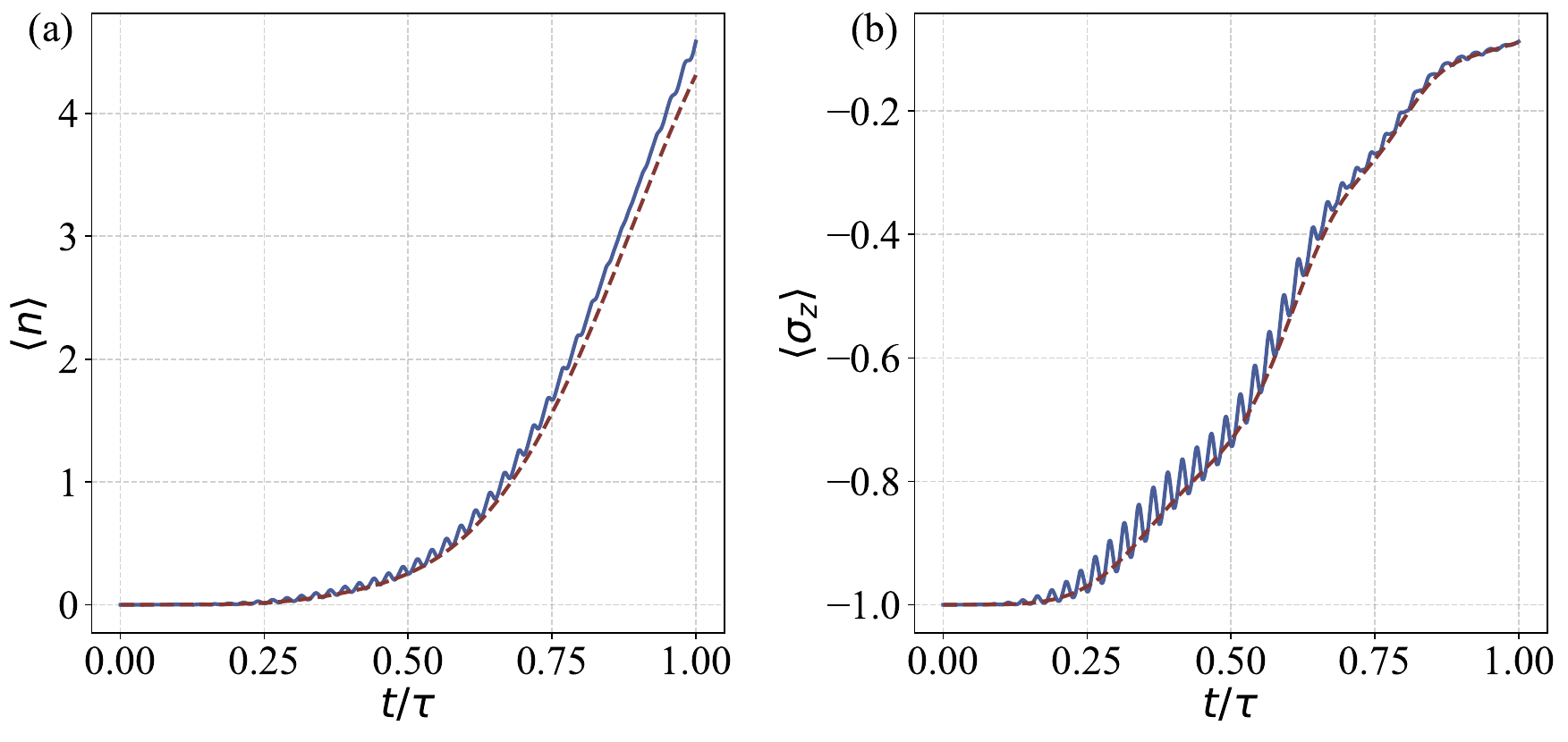}
    \caption{Comparison between exact CD evolution (dashed red) and Floquet-engineered CD dynamics (solid blue) in the QRM  at resonance $(\Gamma=1)$ in the DSC $(\eta = 1.5)$, where (a) time evolution of the cavity excitation number $\langle \hat n\rangle$ and (b) time evolution of the atomic polarization $\langle \hat\sigma_z\rangle$. The simulations are performed for a dimensionless final time $\omega\tau=1$, with Floquet driving frequency $\nu/\omega=40$ and reference frequency scale $\nu_0=1$. }
    \label{fig: floquet_sim}
\end{figure}

We compare the exact CD-assisted evolution with its Floquet-engineered implementation in Fig.~\ref{fig: floquet_sim}. The figure shows the time evolution of the cavity occupation $\langle n \rangle$ [Fig. \ref{fig: floquet_sim}(a)] and the atomic polarization $\langle \sigma_z \rangle$ [Fig.\ref{fig: floquet_sim}(b)] for different total evolution times. The dynamics are driven at resonance in the DSC regime, and the AGP parameters are obtained from the variance-based metric introduced in the previous section. The results demonstrate that the CD trajectory of the Rabi system can be faithfully reproduced without explicitly engineering additional static interaction terms. Instead, the required CD structure emerges dynamically from the high-frequency Floquet modulation. Evaluated at stroboscopic times $t = nT$, the mean fidelity reaches $F= 0.99836$, confirming that Floquet engineering provides an experimentally viable route to implement counterdiabatic control in the quantum Rabi model.

Although parametric modulation requires additional control over the qubit, field, and coupling frequencies, which is not typically available in most experimental realizations, there nevertheless exist platforms capable of supporting its implementation. Several theoretical proposals for Floquet engineered CD protocols have been put forward for systems such as superconducting circuits~\cite{Ribeiro2017,Petiziol2019} and nitrogen vacancy centers where even an experimental realization has also been demonstrated~\cite{Boyers2019}. In the case of superconducting circuits, the high degree of device tunability has been shown to enable large coupling strengths~\cite{FornDiaz2017}. When combined with recent advances in parametric modulation~\cite{Caldwell2018,Jin2025}, this level of control could eventually permit the exploration of the model in the USC and even DSC regimes.

\section{Conclusion}

We have developed and benchmarked a regularized variational CD framework for the QRM  that remains effective across a broad parameter landscape, from the strong- to the deep-strong-coupling regimes and for different detuning conditions. The resulting AGP suppresses diabatic transitions originating from both the atomic and bosonic degrees of freedom, enabling high-fidelity ground-state preparation beyond previous dispersive-regime CD proposals. We also derived an analytical expression for the first-order AGP coefficient, which correctly reproduces the dispersive limit but becomes cutoff dependent outside this regime. This behavior highlights the need for a controlled regularization of the variational metric in unbounded bosonic Hilbert spaces. More broadly, in unbounded Hilbert spaces, the variational metric should be regarded as a physical ingredient of the AGP construction, rather than as a purely numerical choice. This perspective opens a route to designing regularized CD protocols for broader classes of continuous-variable and hybrid quantum systems.

To address the divergences arising from the bosonic mode beyond the dispersive regime, we introduced physically motivated regularization strategies that incorporate prior knowledge of the QRM directly into the variational cost function. We explored three complementary approaches: (i) a weighted trace based on displaced coherent states, (ii) a variance-based metric evaluated with respect to a superradiant-like reference state, and (iii) a filtered trace that restricts the metric to the physically supported Fock-space domain. Across regimes, these approaches regularize the optimization landscape and improve the correlation between the variational action and the resulting ground-state fidelity, thereby enhancing the reliability of AGP optimization. Complementarily, we adopted a fidelity-based quantum optimal-control perspective that bypasses trace-based functionals and enables robust high-fidelity state preparation across the explored Rabi parameter space. Finally, we showed that the resulting CD terms can be implemented via Floquet engineering by parametrically modulating the native couplings of the Hamiltonian, making the protocol compatible with platforms such as nitrogen-vacancy centers and highly tunable superconducting-circuit devices.

\section*{Acknowledgments}
The authors would like to thank J. Casanova, M. Garcia de Andoin and F. Motzoi for useful discussions and insightful comments. The project was partially supported  from the Basque Government through the ELKARTEK program, project "KUBIBIT - Kuantikaren Berrikuntzarako Ikasketa Teknologikoa" (KK-2025/00079) and Newhegaz project (KK-2025/00074). F.A.C.L. thanks to the German Ministry for Education and Research, under QSolid, Grant no. 13N16149 and Horizon Europe program via project QCFD (101080085, HORIZON-CL4-2021-DIGITAL-EMERGING02-10), project OpenSuperQPlus100 (101113946, HORIZON-CL4-2022-QUANTUM-01-SGA). P. G. A. acknowledges support from UPV/EHU Ph.D. Grant No. PIFG 22/25. X.C. appreciates the project grant PID2021-126273NB-I00 funded by MCIN/AEI/10.13039/501100011033 and by ``ERDF A way of making Europe'' and ``ERDF Invest in your Future'', the Severo Ochoa Centres of Excellence program through Grant CEX2024-001445-S, the Spanish Ministry of Economic Affairs and Digital Transformation through the QUANTUM ENIA project call-Quantum Spain project. 

\bibliography{bibl}
\onecolumngrid
\appendix

\clearpage

\section{Analytic derivation of AGP}
\label{Appendix_analytical_trace}

In this Appendix, we derive the analytic expression for the AGP coefficient at first order in the nested-commutator expansion.
To this end, we combine the AGP defined in Eq. (\ref{rabi_AGP}) with the action formalism introduced in Eqs. (\ref{action}) and (\ref{G_operator}).
First, we evaluate the operator $\hat{G}(t;\vec{x})$, which reads
\begin{equation}
    \hat{G}(t;\vec{x}) = \hat{\sigma}_x (\hat{a}^\dagger+\hat{a})+x_1(t) (\Gamma^2+1) \hat{\sigma}_x (\hat{a}^\dagger+\hat{a}) + 2i x_1(t) \Gamma \hat{\sigma}_y (\hat{a}^\dagger-\hat{a}) + 2 \lambda(t)\eta x_1(t) - 2\lambda(t) \eta \Gamma x_1(t) \hat{\sigma}_z (\hat{a}^\dagger+\hat{a})^2.
\end{equation}
Substituting this expression into the trace action gives
\begin{multline}
S(t;\vec{x}) = \bigl(1 + 2 x_1(t)(\Gamma^2 + 1) + x_1(t)^2(\Gamma^2 + 1)^2\bigr)
\, \mathrm{Tr}\!\left[\mathbf{1}_{2\times 2} \otimes (\hat{a}^\dagger + \hat{a})^2\right] 
-\, 4 x_1(t)^2 \Gamma^2 \, \mathrm{Tr}\!\left[\mathbf{1}_{2\times 2} \otimes (\hat{a}^\dagger - \hat{a})^2\right] \\
+\, 4\lambda(t)^2 \eta^2 x_1(t)^2 \, \mathrm{Tr}\!\left[\mathbf{1}_{2\times 2} \otimes \mathbf{1}_{N\times N}\right] 
+\, 4\lambda(t)^2 \eta^2 \Gamma^2 x_1(t)^2 \, \mathrm{Tr}\!\left[\mathbf{1}_{2\times 2} \otimes (\hat{a}^\dagger + \hat{a})^4\right].
\end{multline}
Using $\mathrm{Tr}(\hat A\otimes\hat B)=\mathrm{Tr}(\hat A)\mathrm{Tr}(\hat B)$,  along with, 
\begin{subequations}
\label{eq:rabi_agp_derivation}
\begin{align}
\mathrm{Tr} \left( (\hat{a}^\dagger+\hat{a})^2 \right) &= n(n+1), \\[6pt]
\mathrm{Tr} \left( (\hat{a}^\dagger-\hat{a})^2 \right) &= -n(n+1), \\[6pt]
\mathrm{Tr} \left(\mathbf{1}_{N\times N}\right) &= n+1, \\[6pt]
\mathrm{Tr} \left( (\hat{a}^\dagger+\hat{a})^4 \right) &= n(2n^2+n-1),
\end{align}
\end{subequations}
the action can be evaluated explicitly as a function of the Rabi parameters and the Fock-space cutoff. 
Here, $n$ denotes the Fock-state cutoff, and $N=n+1$ the dimension of the Hilbert space. Taking the derivative of the action with respect to $x_1(t)$ yields
\begin{multline}
\partial S(t;\vec{x})/\partial x_1(t) = \bigl( 4(\Gamma^2 + 1) + 4x_1(t)(\Gamma^2 + 1)^2+16x_1(t) \Gamma^2\bigr)n(n+1) \\
+\, 16\lambda(t)^2 \eta^2 x_1(t) (n+1)
+\, 16\lambda(t)^2 \eta^2 \Gamma^2 x_1(t) n(2n^2+n-1).
\end{multline}
By setting $\partial S/\partial x_1(t)=0$, and solving the equation, we obtain the analytical coefficient that minimizes the action.
\begin{equation}
    x_1(t) = -\frac{1+\Gamma^2}{1+\Gamma^4+6\Gamma^2+4\lambda(t)^2\eta^2(\frac{1}{n}+(2n-1)\Gamma^2)}.
\end{equation}
This expression reveals that, in addition to the parameters explicitly appearing in the quantum Rabi Hamiltonian, the AGP coefficient also depends on the truncation of the bosonic Hilbert space. In the large-cutoff limit, $n\to\infty$, the coefficient vanishes for finite $\Gamma$, reflecting the dimensionality problems of the trace action in an unbounded bosonic Hilbert space. In contrast, in the dispersive limit $\Gamma\to0$, the cutoff-dependent contribution disappears and the expression recovers the known dispersive-regime CD correction~\cite{Nori_sta}. This result provides both an analytical benchmark for the variational construction and a clear diagnosis of the divergence problem addressed in the main text.


\section{Renormalization trace schemes: Optimization landscapes and Correlations}
\label{app:variational_trace}

We analyze how the choice of reference operator affects the variational action associated with the QRM.   Our goal is to show that regularization does more than reduce the effective dimension of the search space: it also modifies the topology and gradients of the optimization landscape, thereby improving the relation between the minimized action and the final state-preparation fidelity.

\begin{figure}[t]
    \centering
    \includegraphics[width=\textwidth]{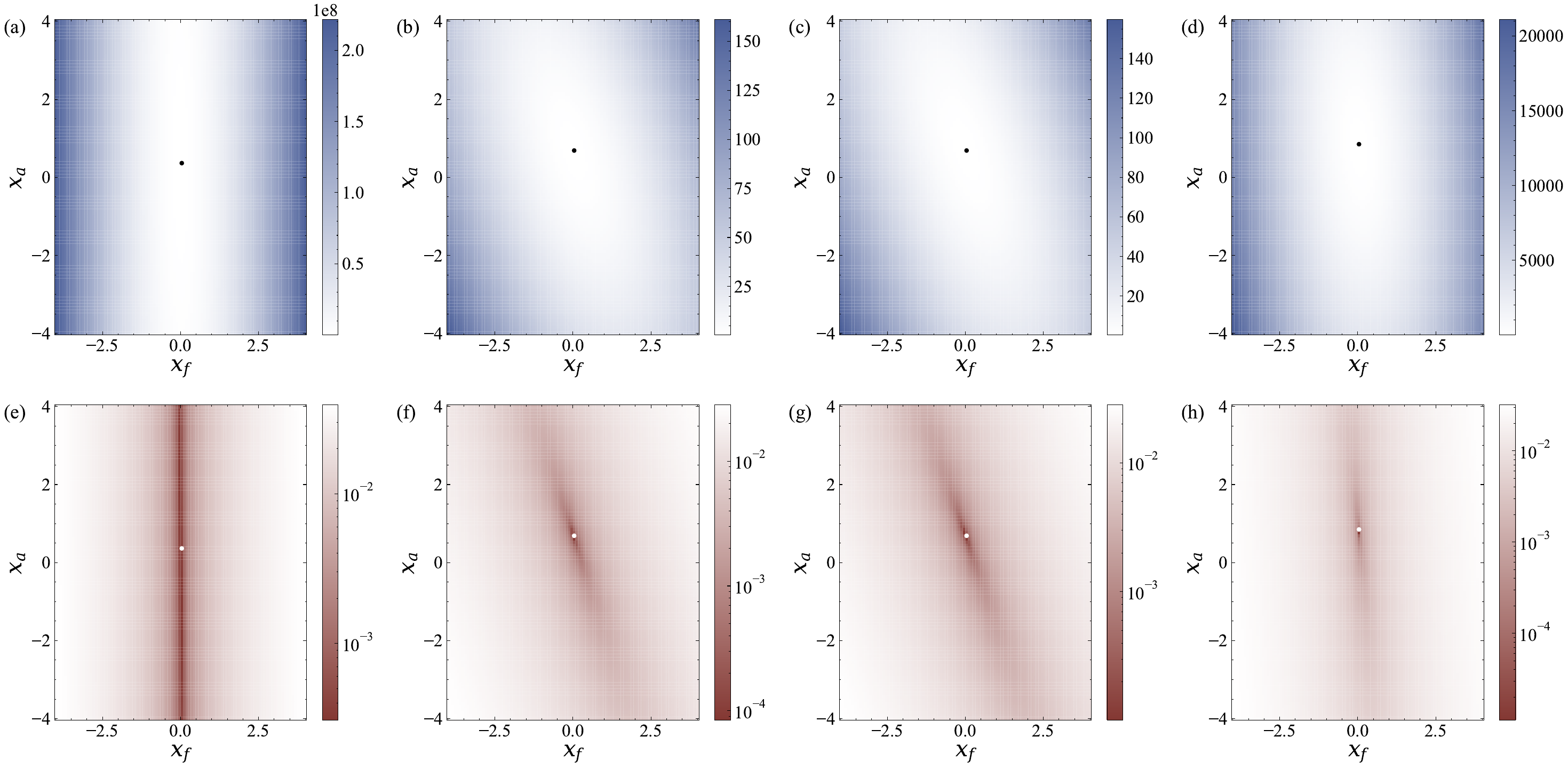}
    \caption{Action landscape and corresponding gradients for different variational CD cost functionals, evaluated as functions of the AGP coefficients. The figure shows both the action value and its gradient at resonance in the USC regime, evaluated at time $t=0.5\tau$.  Panels are grouped as (a,e) standard trace action, (b,f) coherent-state weighted trace, (c,g) variance-based superradiant metric, and (d,h) filtered-trace functional. The results show that the regularized metrics not only reduce the overall magnitude of the action but also reshape the optimization landscape, suppressing barren-plateau-like regions and restoring useful gradient information.}
    \label{gradient_trace_manifolds}
\end{figure}

To this end, we compute the action values over a range of AGP parameters for the QRM at resonance in the USC regime. The top panels of Fig. \ref{gradient_trace_manifolds} show that the regularized metrics effectively stabilize the magnitude of the action and suppress the cutoff-induced growth associated with the bosonic Hilbert-space truncation. The effect of regularization, however, goes beyond dimensional reduction. 
The lower panels of Fig.~\ref{gradient_trace_manifolds} show the corresponding gradient landscapes, computed using second-order central finite differences. These results demonstrate that the reference states reshape the gradient structure of the action functional. In particular, the unregularized trace action exhibits a barren-plateau-like region in the atomic component [Fig.~\ref{gradient_trace_manifolds}(e)], whereas the regularized metrics restore finite gradients and provide a more favorable landscape for minimization.

\begin{figure}[h]
    \centering
    \includegraphics[width=\textwidth]{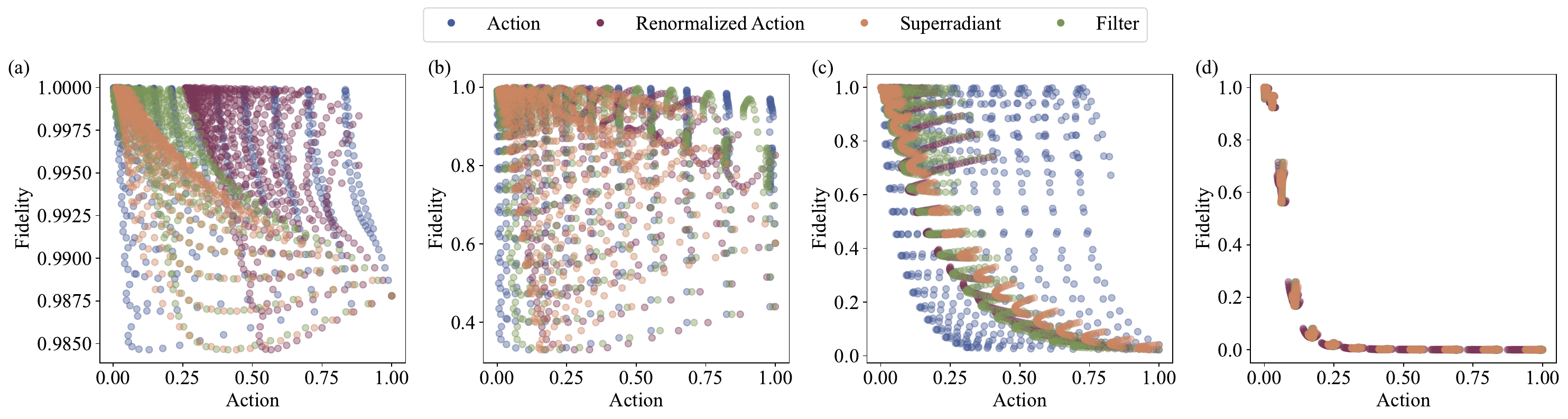}
    \caption{Comparison between the accumulated normalized action and the final ground-state fidelity for different regularized metrics. The panels correspond to (a) the UC regime $\left(\omega,\Gamma,\eta\right) = \left(1,1,0.25\right)$, (b) the USC regime $\left(1,1,0.25\right)$, (c) the high-cavity-frequency regime $\left(1,0.1,1\right)$, and (d) the high-atom-frequency regime $\left(1,10,1\right)$. In each panel, we compare the standard trace action, the coherent-state weighted trace, the variance-based superradiant metric, and the filtered-trace protocol for different AGP coefficients $\vec{x}$. The results show how physically motivated regularization improves the correlation between action minimization and state-preparation fidelity across distinct light--matter regimes.}
    \label{fig: trace_vs_fidelity}
\end{figure}

Finally, we investigate the reliability of the action as a figure of merit for optimizing the AGP coefficients in unbounded systems, and examine how regularized metrics improve its predictive power. To this end, we analyze the statistical correlation between the final ground-state fidelity and the accumulated action, defined as 
\begin{equation*}
   S =  \int_0^1 \norm{F_k(\vec{x};\lambda)} d \lambda.
\end{equation*}
where $F_k(\vec{x};\lambda)$ denotes one of the action functionals defined in the main text and $\vec{x}$ represents the AGP coefficients. Fig.~\ref{fig: trace_vs_fidelity} illustrates the relationship between the final ground-state fidelity and the normalized accumulated action across different resonance and interaction regimes with different AGP parameters $\vec{x}$. We compare the standard trace action with three alternative schemes: the coherent-state weighted trace, the variance-based superradiant metric, and the filtered-trace functional. In the SC, USC, and high-cavity-frequency regimes [Figs.~\ref{fig: trace_vs_fidelity}(a)--\ref{fig: trace_vs_fidelity}(c)], the regularized metrics substantially enhance the correlation between the action and the final fidelity, indicating that restricting the metric to a physically relevant subspace improves its predictive capability. By contrast, in the atom-frequency-dominated regime [Fig.~\ref{fig: trace_vs_fidelity}(d)], the improvement is weaker, indicating limitations of these reference-state choices in that regime.

To quantify the correlation between the accumulated action and the final fidelity, we compute Spearman's rank correlation coefficient,
\begin{equation*}
    r_s  = \frac{{\rm{cov}}[R[X],R[Y]]}{\sigma_{R[X]}\sigma_{R[y]}},
\end{equation*}
where  $R[X]$ and $R[Y]$ denotes the rank variables associated with the data sets X and Y,  $cov[R[X],R[Y]]$ is their covariance, and $\sigma_{R[X]}$ and $\sigma_{R[Y]}$ are their standard deviation. The Spearman correlation coefficient takes values in the range  of $r_s \in [-1,1]$,  quantifying the strength and direction of the monotonic correlation. The values of $r_s$ obtained from our analysis are summarized in the Table \ref{spearsman_table} (Sec. \ref{sec:renormalized_CD}), providing a quantitative comparison of the predictive power of the different action-based metrics across the considered regimes.

\section{Floquet engieneering of the adiabatic gauge potential}
\label{floquet}

Floquet theory provides a powerful framework to treat periodic differential equations of the form
\begin{equation}
    \dot{x}(t) = A(t) x(t), \quad A(t) = A(t+T).
\end{equation}
The Floquet's theorem guarantees that the solution of $x(t)$ can be decomposed into a product of a periodic component $P(t)=P(t+T)$ and an the time-independent exponential factor $F$
\begin{equation}
    x(t) = P(t) \exp{(tF)}.
\end{equation}
The Floquet theory can be translated into the quantum mechanics formalism \cite{universal_floq,floquet_qs}, here the evolution of a periodically driven Hamiltonian $\hat{\mathcal{H}}_E(t) = \hat{\mathcal{H}}_E(t+T)$ can be expressed over one period via the time-independent Floquet Hamiltonian $\hat{\mathcal{H}}_F$
\begin{equation}
    \exp \left( -i \hat{\mathcal{H}}_F T\right) = \mathcal{T} \exp \left( -i \int_t^{t+T} \hat{\mathcal{H}}_E(\tau) d\tau \right), 
\end{equation}
with $\mathcal{T}$ the time-ordering operator. The complete instantaneous Hamiltonian can be written as
\begin{equation}
    \hat{\mathcal{H}_E}(t) = P(t) \hat{\mathcal{H}}_F P^\dagger(t) + i \partial_t P(t)P^\dagger(t), 
\end{equation}
where $P(t) = \exp{-i\mathcal{K}(t)}$ the micromotion operator generated by the kick operator $\mathcal{K}(t)$. Here we also set $\hbar = 1 $. In general, obtanining the Floquet Hamiltonian exactly is not feassible. However in the high-frequency regime $$  \nu \gg \{\omega,\Omega,g\},$$ with $\nu = \frac{2\pi}{T}$, the Floquet Hamiltonian can be constructed perturbatively using the Magnus expansion,
\begin{equation}
    \label{Magnus_floquet}
    \hat{\mathcal{H}}_F = \frac{1}{T} \int_0^T \hat{\mathcal{H}}_E(t) dt + \frac{1}{2T} \int_0^T dt \int_0^{t'} \left[\hat{\mathcal{H}}_E(t),\hat{\mathcal{H}}_E(t')\right] dt' + \mathcal{O}(\nu^{-2}).
\end{equation}
The nested commutator structure of Eq. (\ref{Magnus_floquet}) naturally mirrors the structure of the AGP, making Floquet engineering a direct route to implement counterdiabatic driving.  We consider a periodically modulated Hamiltonian for the Rabi problem
\begin{equation}
    \hat{\mathcal{H}}_E(t) =   \hat{\mathcal{H}}_R(t) + \frac{\nu}{\nu_0} \left[ A_c(t) \cos(\nu t)  \hat{a}^\dagger\hat{a} +   A_a(t) \cos(\nu t)  \hat{\sigma}_z\right]  +  \dot{\lambda} \sum_{k=1}^\infty \beta_k \sin((2k-1)\nu t) \eta\hat{\sigma}_x (\hat{a}^\dagger+ \hat{a}),
\end{equation}
where $\hat{H}_R(t)$ is the time-dependent Rabi Hamiltonian, $A_c(t)$ and $A_a(t)$ are the slow varying envelope amplitudes, $\nu_0$ is the renormalization frequency, $\beta_k$ are the Fourier coefficients and $\eta$ is the renormalized coupling constant. By deriving the Magnus equation for the oscilating Hamiltonian we get for the leading term 
\begin{equation}
    \hat{\mathcal{H}}_F^{(0)} = \frac{1}{T} \int_0^T \hat{\mathcal{H}}_R(t) dt,
\end{equation}
which captures the average dynamics of the bare Hamiltonian. In the first order of correction, only the cross term contributions survives and the Hamiltonian can be written as
\begin{equation}
\label{first_order_ME}
    \hat{\mathcal{H}}_F^{(1)} = \frac{\eta }{2T} \frac{\nu}{\nu_0} \left(\bar{A}_c\left[\hat{a}^\dagger \hat{a},\hat{\sigma}_x (\hat{a}^\dagger+\hat{a})\right] + \bar{A}_a\left[\hat{\sigma}_z,\hat{\sigma}_x (\hat{a}^\dagger+\hat{a})\right] \right) \int_0^T  \int_0^{t'} \cos{(\nu t)} \sin{((2k-1)\nu t')}dt dt',
\end{equation}
where $\bar{A}_{c/a}$ are the envelope-averaged amplitudes, and the integral is
\begin{equation}
\label{integral_ME}
    I_k =\int_0^T  \int_0^{t'} \cos{(\nu t)} \sin{((2k-1)\nu t')}dt dt'= \begin{cases}
    -\frac{\pi}{\nu^2} \quad  k = 1 \\
   0 \quad  k>1
    \end{cases}.
\end{equation}
Thus, only the first harmonic contributes at leading order, justifying truncation to $k=1$. By substituding the commutators in the Magnus expression one get
\begin{equation}
      \hat{\mathcal{H}}_F^{(1)} = \dot{\lambda} \left[ \frac{\beta \eta \bar{A}_c}{4 \nu_0} i \hat{\sigma}_x (\hat{a}^\dagger-\hat{a}) + \frac{\beta \eta \bar{A}_a}{2 \nu_0} \hat{\sigma}_y (\hat{a}^\dagger+\hat{a})\right].
\end{equation}
This reproduces exactly the operator structure of the counterdiabatic Hamiltonian derived from the AGP. Higher-order Magnus terms generate more expansion of the AGP, providing a systematic route to engineer increasingly accurate AGP approximations. These contributions are suppressed by higher powers of $\nu$. In higher orders, the integral $I_k$ is not longer zero for $k> 1$, and the integral can expressed in terms of Bessel functions $\mathcal{J}_k(\nu/\nu_0)$, leading to a more complex control scenario.

\end{document}